# Single atom manipulation and control in a scanning transmission electron microscope

# Single atom manipulation and control in a scanning transmission electron microscope

Ondrej Dyck[1], Songkil Kim[1], Sergei V. Kalinin[1], and Stephen Jesse[1]

[1]The Institute for Functional Imaging of Materials and The Center for Nanophase materials sciences, Oak Ridge National Laboratory, Oak Ridge, Tennessee, 37831, USA

We demonstrate that the sub-atomically focused beam of a scanning transmission electron microscope (STEM) can be used to controllably manipulate individual dopant atoms in a 2D graphene lattice. We demonstrate the manipulation of adsorbed source materials and the graphene lattice with the electron beam such that individual vacancy defects can be controllably passivated by Si substitutional atoms. We further demonstrate that these Si defects may be directed through the lattice via e-beam control or modified (as yet, uncontrollably) to form new defects which can incorporate new atoms into the graphene lattice. These studies demonstrate the potential of STEM for atom-by-atom nanofabrication and fundamental studies of chemical reactions in 2D materials on the atomic level.

Fabrication of structure atom-by-atom has remained one of the longest-held dreams of nanoscience, as a key element of nanotechnology and penultimate step for understanding physics and chemistry on the atomic level. The development of scanning tunneling microscopy (STM) in the early 1980s has demonstrated the potential of an atomically sharp tip to induce atomic motion on a surface, originally perceived to be detrimental to microscope operation. However, the work by Eigler at IBM in the early 1980s demonstrated that tip induced atomic motion can be used for the assembly of functional atomic structures,[1-5] an accomplishment believed to be one of the key factors that lead to the nanotechnology revolution of the last decades. However, STM operation necessitates low temperature ultra-high vacuum environments and typically results in structures confined to reactive surfaces. Correspondingly, it took over 20 years to transition from atomic manipulation by STM to viable pathways for single-atom devices, as demonstrated by M. Simmons and others.[6-8] This, in turn, necessitates the search for alternative methods for single atom manipulation and atom-by-atom assembly.

By necessity, such methods must combine both the imaging and manipulation steps, to affect matter on the atomic level and to observe initial, intermediate, and final stages of the system on the atomic level. Beyond scanning probe microscopies, the natural candidate for this is the scanning transmission electron microscope (STEM), based on the atomically-focused electron probe, the position of which can be controlled with picometer precision.[9-11] Since the early days of atomically resolved STEM, it has been realized that electron beams can strongly affect the crystalline lattice, leading to a broad range of transformations typically associated with damage and degradation of the material.[12, 13] However, in the last several years it has been shown that the e-beam can introduce much more subtle changes in a material's structure that can be resolved on the atomic level,[14-21] including vacancy ordering,[19, 22-26] single dopant atom motion in 2D and 3D materials,[27-31] and chemical reactions.[32, 33]

Here, we explore the potential of the atomically focused e-beam to manipulate single Si atoms on a graphene lattice, including directed motion and incorporation in the lattice. As a key enabling component, we utilize direct e-beam control.[27, 34-38]

As a model system, we have chosen CVD-grown graphene, transferred from the Cu foil growth substrate to a TEM sample grid followed by a Ar-$O_2$ anneal at 500 °C for removal of volatile adsorbents. The Cu foil was spin-coated with PMMA to stabilize the graphene and the Cu foil was etched away in a bath of ammonium persulfate-DI water solution. The graphene/PMMA layer was transferred to a DI water bath to remove residues of ammonium persulfate. The graphene was transferred to the final TEM substrate by scooping it from the bath and letting it dry at room temperature. Following the recipe of Garcia et. al.,[39] TEM samples were baked in an oven under an Ar-O$_2$ environment to remove residual PMMA and volatile organic compounds. Imaging of the samples was performed in a Nion UltraSTEM 100 at an accelerating voltage of 100 kV and 60 kV in either high angle annular dark field (HAADF) or medium angle annular dark field (MAADF) imaging modes, as indicated in the text.

As a first step toward demonstrating atomic-level control of single atoms, single impurity atoms must be introduced into the graphene lattice. While obtaining impurity atoms on the surface of graphene is rather trivial (obtaining clean graphene is the real challenge), introducing them into the graphene lattice in a controllable way at the atomic scale is less obvious. To this end, we note that the amorphous source/contaminant material, comprised of mostly amorphous carbon and silicon atoms, is readily sputtered away with a 100 keV beam and Si atoms are scattered across the surface of the graphene. Since the beam has a higher energy (at 100 keV) than the knock-on energy for graphene, defects are concurrently created in the graphene, and silicon substitutional atoms are incorporated randomly into the lattice. The Si atoms that are freed from the source but have not bonded to the graphene lattice show up in the image as evanescent streaks as they move readily under the beam.

Progressing toward more precise control of this process, we demonstrate introduction of a single Si substitutional defect at a specific lattice site (instead of randomly). To achieve this, we made use of graphene's ability to self-heal.[14, 40-42] The 100 keV STEM beam was placed on the desired lattice site, which was briefly (~1-2 s) exposed to the beam, to create a defect site, FIG. 1 a). Once a defect was created, b), a small (~1-2 nm) scan area was selected over the Si/C source material and sputtered away from the beam onto the graphene lattice, shown in c). Since the source material is composed of mainly amorphous carbon and silicon atoms, and since graphene tends to self-heal, there is a high likelihood of the lattice healing by incorporating the source atoms. Thus, a single Si atom is introduced into the graphene lattice.

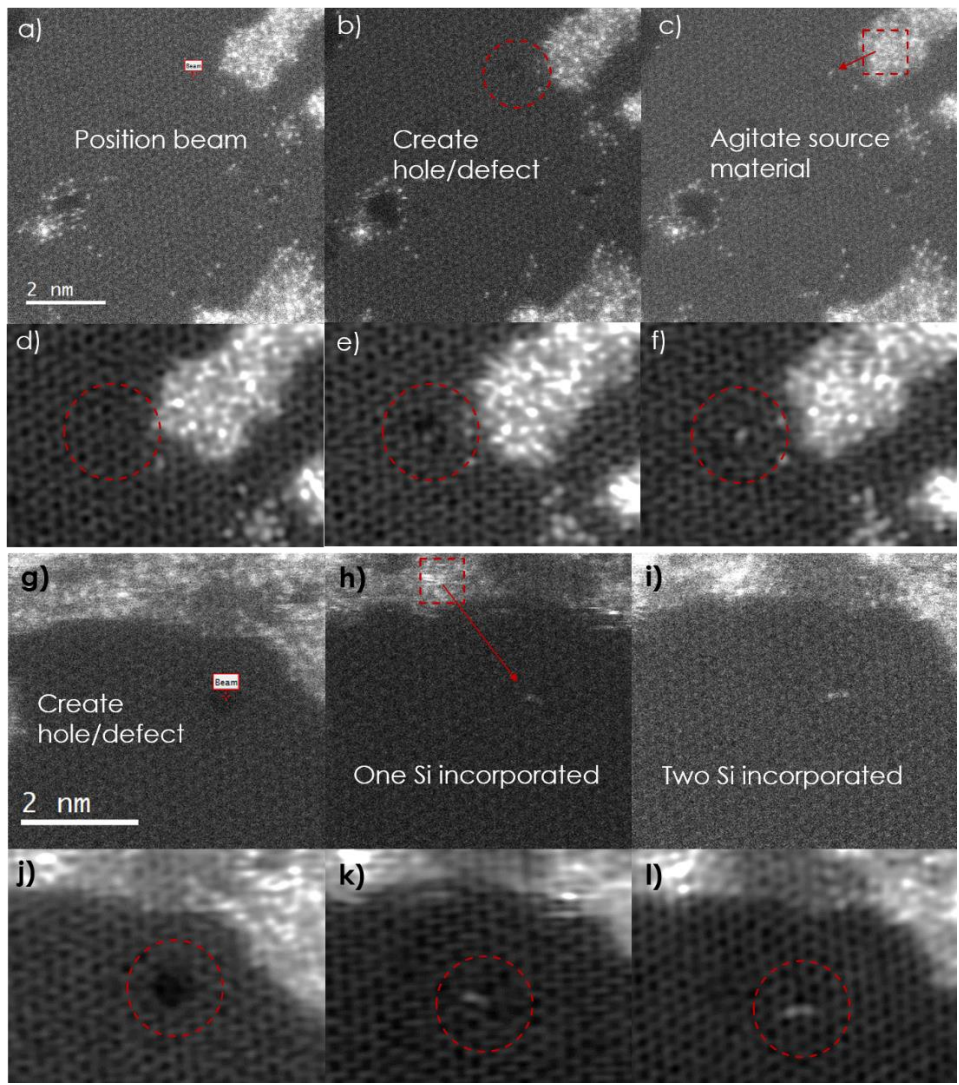

FIG. 1. Illustration of the introduction of a single Si substitutional defect. a) shows the positioning of the electron beam over the desired location. The beam is briefly un-blanked to create a small hole or defect in the graphene lattice, b). Once the defect has been created, a small sub-scan is performed over the source material to sputter the carbon and silicon atoms into the defect. The graphene lattice spontaneously heals using the source atoms. d)-f) are cropped from a)-c) and filtered to provide a clearer view of the graphene lattice at each step. A further example of introducing Si defects in a graphene lattice is shown in g)-l). We first create a hole or defect, shown in the MAADF image in g). Second, we performed a small sub-scan over the source material to scatter the desired atoms into the defect site, shown in h). Finally, the graphene lattice healed, incorporating a Si dimer, shown in i). j)-l) were cropped from g)-i) and filtered to more clearly show the graphene lattice at each stage.

FIG. 1 d)-f) show the region of interest cropped from a)-c) and filtered to more clearly show the graphene lattice during the above described procedure. Limiting beam exposure, particularly after the creation of a hole is important to prevent continued hole growth or introduction of new defects within the area of interest.

As a further demonstration, we were also able to introduce silicon dimers into the graphene lattice as well. FIG. 1 g)-l) show an example. Using the same procedure, we first created a small hole, g), then performed a sub-scan over a small area of source material and allowed the graphene lattice to heal with the scattered source atoms, shown in h). Finally, the graphene fully healed, incorporating a silicon dimer, shown in i). FIG. 1 j)-l) were cropped from g)-i), respectively, and filtered to show the lattice more clearly at each stage.

Controllable motion of single silicon substitutional defects in graphene was also accomplished, shown in FIG. 2. A Si atom's motion through a graphene lattice is caused by the temporary removal of a neighboring carbon atom when absorbing energy from the beam. The Si atom moves over to fill the vacancy while simultaneously pulling the ejected carbon atom into the lattice behind it. This process is detailed by Susi et. al..[27] To instigate directed motion of the silicon atom, a small scan was performed with a 60kV electron beam over the Si defect and the nearest neighbor carbon atoms in the desired direction of motion, FIG. 2 g) and h).

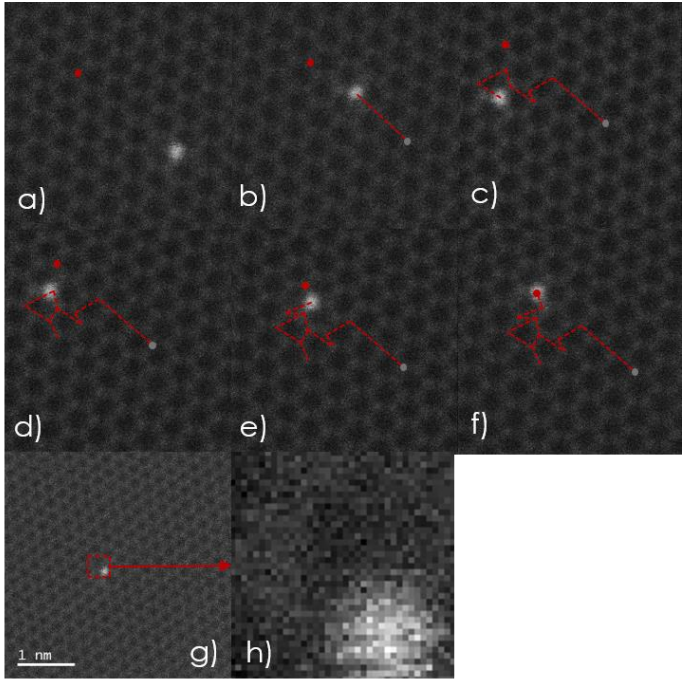

FIG. 2. (color online) Example of moving a Si atom through the graphene lattice. a)-f) shows the progression through time of a Si atom pushed through a graphene lattice with a 60kV electron beam. The gray dot in b)-f) marks the original location of the Si atom from a). The red dot marks the target lattice site (where we were trying to get the Si atom to move to). The dotted line tracks the position of the Si atom through the lattice for each acquired image. The full image dataset may be found in the supplementary information. Note: between d) and e) the image acquisition area was shifted to prevent the target lattice site from drifting out of the field of view. g) and h) illustrate the method used to move the Si atom in a)-f). A small sub-scan area is irradiated (boxed in g)) including the silicon atom and the neighboring carbons in the direction of desired movement. One frame of the sub-scan used is shown in h).

Though Si atoms typically move at random under the beam, this procedure increases the probability of transition toward the sub-scan area. A similar procedure was used by Susi et. al.[43] to achieve controlled motion of a Si atom in graphene.

In addition, we also observed defect evolution under the prodding of the 60kV electron beam. FIG. 3 summarizes the observed evolution of two 4-fold coordinated Si defects, both of which began by the removal of an adjacent carbon atom. FIG. 3 a)-d) shows the evolution of the first example. a) shows a high angle annular dark field (HAADF) image of the initial state. We note that a 4-fold coordinated Si defect replaces two carbon atoms from the lattice.

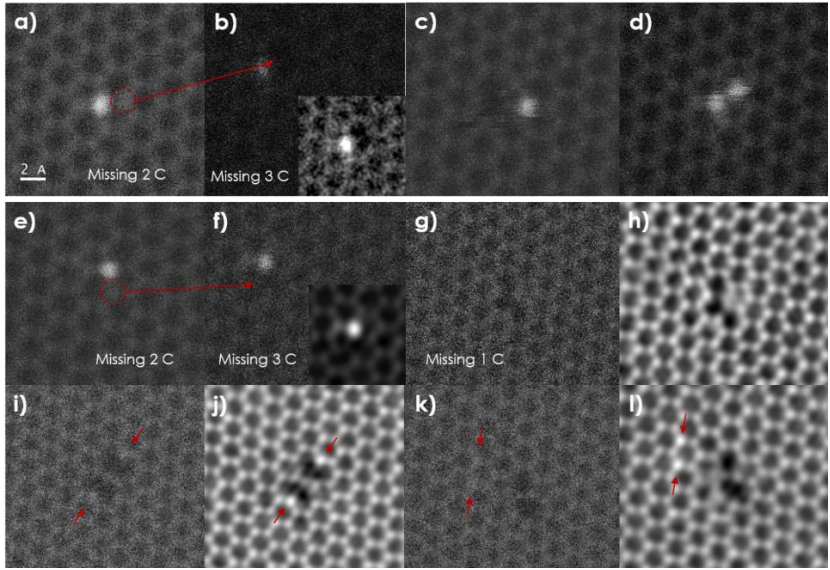

FIG. 3. Evolution of point defects. a)-d) show the evolution of a point defect under 60kV electron beam irradiation. First a carbon atom is ejected from a lattice location adjacent to a substitutional silicon (indicated by the circle and arrow in a) and b)). The inset in b) shows a smoothed and brightened version of b) to more clearly show the defect. This state was fleeting which prevented acquisition of a higher quality image. The image shown in c) is the result of acquiring a higher quality image where we observe the graphene lattice restructuring during image acquisition (atoms appear torn around the defect indicating rapid motion). Finally, the lattice stabilizes by acquiring a second substitutional Si, shown in d). e)-l) show the evolution of another point defect. A carbon atom adjacent to the silicon is ejected from the lattice through e-beam irradiation, indicated by the circle and arrow in e and f). The inset in f) shows a filtered of the original image to more clearly show the vacancy. After continued irradiation the silicon atom appears to be replaced by two carbon atoms, shown in g). h) shows a filtered version of g). Further irradiation inspires another reconfiguration incorporating two brighter atoms indicated by the arrows in i). j) shows a filtered version of i). Finally, the defect restructures again, incorporating the two brighter atoms into the lattice, indicated by arrows in k). l) is a filtered version of k).

Under the scanning of the 60kV electron beam a C atom adjacent to the Si is knocked from the lattice, shown in b). The image shown in b) was obtained during a quick scan normally used for rapid sample viewing and not final image acquisition which resulted in low a low-quality image. The inset shows a smoothed and brightness/contrast adjusted version to more clearly show the C vacancy next to the Si. Attempting to acquire a high-quality image of the defect resulted in the image shown in c), where we observe that the defect was restructuring under the beam irradiation, which produced streaking of the atoms around the defect site. Finally, the lattice healed by incorporating a second Si atom, forming a clover-like shape comprising three 5-fold rings with the two Si atoms at the core, shown in d).

A second example of beam-induced defect evolution is shown in e)-l). The defect began with a 4-fold coordinated Si as shown in e). The 60kV electron beam was scanned over an adjacent carbon, ejecting it from the lattice, f). The inset in f) shows a filtered version of the image to more clearly show the defect. Upon additional disturbance from the electron beam (normal scanning), the Si atom was ejected from the lattice, shown in g). A filtered version of g), shown in h), indicates that the lattice is no longer missing three carbon atoms. It appears that while the Si atom was ejected, two carbon atoms were acquired. We posit that there is still one C atom missing and a single C atom is quickly oscillating between two positions at the center of the defect. This defect appears to most closely resemble the saddle point described by El-Barbary et. al.[22] addressing vacancy movement. Nevertheless, this defect configuration was relatively stable, so that several images could be acquired before it reconfigured to the structure shown in i). A filtered version of i) is shown in j). This defect appears to involve two new atoms that are slightly heavier (possibly N), and thus, brighter in HAADF imaging, indicated by arrows. This configuration may be the result of two adjacent single N pyridinic vacancies.[44, 45] Upon continued irradiation from the electron beam, the structure transformed again into that shown in k). A filtered version of this image is shown in l). The heavier/brighter atoms have now been incorporated into the lattice.

These results illustrate that controllable introduction of specific species of dopant atoms, here silicon, into a graphene lattice is possible. Moreover, once a dopant silicon atom has been introduced into the lattice, it is possible to induce directed motion through the lattice with the e-beam. The silicon defect can also be altered in an (as yet) uncontrollable way to produce a variety of other defects which themselves may incorporate foreign atoms into the graphene lattice. Importantly, we introduce the idea of a solid-state source material that may be sputtered into a graphene defect as a method for introducing foreign atoms into the lattice. One can easily imagine introducing a variety of source materials on the carbon lattice and creating different species of defects that may be brought together for atomic scale fabrication through e-beam modification. Such capabilities would prove invaluable toward understanding the behavior of few atom systems which could be built and then characterized, or the fabrication of nanoscale building blocks for molecular machines.

These studies establish new opportunities for exploring beam-induced chemical reactions and properties of dopant atoms in new and unusual coordinations, enabled by highly non-equilibrium environments. Furthermore, this sets the background for atom-by-atom assembly. To accomplish this task, many microscope and sample related problems still need to be solved, including the separation of imaging and manipulation stages e.g. via compressed sensing and AI assisted imaging, feedback loops for automatic atomic motion control, the development of theory to establish mechanisms to achieve reasonable fabrication rates, and sample preparation techniques to allow reproducible mesoscale control of source materials.

**Supplementary Information**

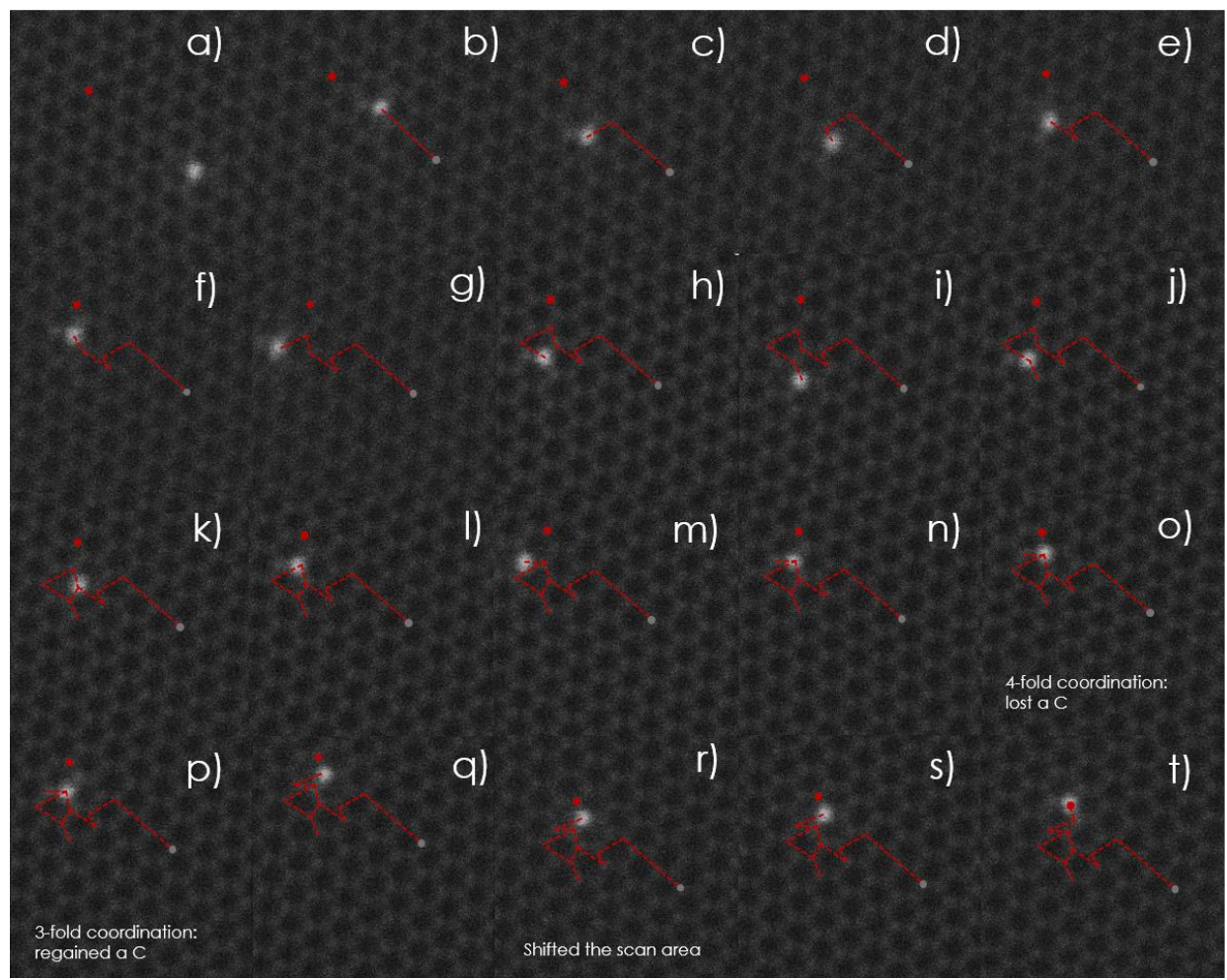

**Figure S1** a)-t) show the progression through time of the Si atom motion. In o) the Si atom switched to 4-fold coordination through loss of an adjacent carbon atom. In p) a carbon atom is regained and the structure switches back to 3-fold coordination. Between q) and r) the scan area was shifted slightly to compensate for a small amount of drift.

**Acknowledgements**

We would like to thank Dr. Ivan Vlassiouk for provision of the graphene samples and Dr. Francois Amet for performing the argon-oxygen cleaning procedure.

Research supported by Oak Ridge National Laboratory's Center for Nanophase Materials Sciences (CNMS), which is sponsored by the Scientific User Facilities Division, Office of Basic Energy Sciences, U.S. Department of Energy (S.V.K.), and by the Laboratory Directed Research and Development Program of Oak Ridge National Laboratory, managed by UT-Battelle, LLC, for the U.S. Department of Energy (O.D, S.K.,S.J.).